# 氧化石墨烯吸附增强应用的展望

孙 凌[1,2]

1. 北京古月新材料研究院，北京工业大学
2. 新材料与产业技术北京研究院

Email: sunling@bjut.edu.cn

**摘要**： 使用吸附剂是水体修复重要手段，碳是一类重要的吸附剂材料。不同形态的碳材料展现出不同的污染吸附去除能力。氧化石墨烯作为二维碳材料石墨烯衍生的氧化物，继承后者的二维形貌，但结构上嫁接有丰富且差异分布的多种含氧基团。在水中自由舒展的氧化石墨烯类于平面超级大分子，布朗运动作用下会大量接触吸附质，发生强大吸附作用；但含氧基团吸附机制差异、分布差异造成石墨烯表面吸附物质分布相对不均匀。因此，调控表面含氧基团及其分布是吸附增强的重要手段，由此演绎出来的原位基团调控致吸附增强方法是十分有潜力的。

# Graphene oxide adsorptive power from better to more via an enhanced route in perspective

Sun, Ling[1,2]

1. Beijing Guyue New Materials Research Institute, Beijing University of Technology, Beijing
2. Materials and Industrial Technology Research Institute, Beijing

**Abstract**: Adsorption is one important way applied to water decontamination, where carbon is commonly used as highly effective absorbent. Carbon of different morphologies and structures normally demonstrate distinct capabilities to adsorption-typed decontaminations. Graphene oxide is regarded as the oxidative derivative of graphene, maintaining its 2D morphology-featuring structure while location-dependently immobilized with a large amount of various oxygen-involving functional moieties on both sides. Free-standing graphene oxide tends to fully unfold in water and behave in macromolecular-like Brownian motion. Therefore, high-performance adsorption occurs during the process of interacting with large amount of sorbates, during which an unevenness of effective adsorptive site distribution turns out for targeted matter due to functional group-to-sorbate interaction mechanism differentiation as well as groups' inherent location specifications. Thus, oriented manipulation of surficial functional moieties and their site distribution is potentially a way to further enhance the adsorptive performances, and this could go even better if considering combining adsorption and an in-situ manipulation at the same time.

全球范围内，尤其是发展中、不发达国家，水与土壤等环境污染问题比较突出，**环境修复需求迫切**。氧化石墨烯是环境修复材料领域中的研究热点，是极具发展潜力的碳吸附材料。以氧化石墨烯为吸附材料的研究论文数量（WOS 统计截止到 2017/2/28），中国第一（占据全球总量的 63%），紧跟之后的是美国（7.9%）、印度（7.7%）、韩国（7.5%）；自 2012 起全球每年增幅可达 5~7%；论文数量分布存在地区性差异，跟地区环境问题正关联。我国在氧化石墨烯研究的现状，推测主要有两点原因：一是我国部分地区长期奉行"重经济发展轻环境保护"等隐形发展策略，导致生态环境恶化（大气雾霾、源头水质恶化、土壤毒化等），影响民生，给社会发展带来了严重的负面效应；同时，也严重不利于维持国家经济中长期中高水平发展所需人才引进和技术开发氛围的营造。因此，随着社会渴求清洁环境回归，百姓环保意识增强，加之全球对温室气体排放限制、低碳社会发展要求，使政府日益关注先进环境修复材料的研究和开发。在相关政策推动下，各类研究机构依托自身优势积极响应和发力。二是我国与其他国家相比，石墨矿产资源丰富，原材料廉价易得、制备方法比较简单，基于氧化石墨烯的基础和应用研究容易开展。值得注意的是，作为发展中国家的印度，该领域研究论文年增速高达 15.8%，远高于中国的 3.9%，从侧面反映出该国对新型碳材料前沿技术的关注，和用碳纳米技术解决其自身环境问题的渴望。

**氧化石墨烯结构独特、性能优异**。氧化石墨烯作为石墨烯的氧化衍生物，具有碳蜂窝状结构和独特的二维层状形貌。石墨烯仅单原子厚度，比表面积理论上达到 2630 $m^2/g$[1]。用强碱活化石墨烯，其孔结构增多，比表面积可增至 3500 $m^2/g$ 以上，且表现出极优异的超电容和储氢性能[2-4]。利用石墨或石墨烯氧化制备氧化石墨烯，部分碳原子会被氧化、碳结构丧失，仅原子厚度的碳结构上会嫁接有大量氧基团。氧含量差异影响着单层氧化石墨烯的厚度，一般在 0.4~1.1 nm 之间，略厚于石墨烯（~0.34 nm），故氧化石墨烯理应具有高比表面积。实际上，因层间作用存在（范德华力、氢键等）使得结构堆叠、类石墨化，因此文献中报道的比表面积值（多数通过氮吸脱附测定）多在几百 $m^2/g$，远小于石墨烯[5, 6]。但在水中，氧化石墨烯表面大量的亲水氧基团与水分子作用，使石墨烯片层充分单分散开，此时氧化石墨烯的表面结构充分暴露。作为新型非金属型碳催化剂，氧化石墨烯可在水中促进醇、炔、烯烃结构氧化或水合作用[7]；作为新型吸附剂，氧化石墨烯水溶液可直接用于对各类污染靶向物（譬如重金属离子、难降解染料分子等）的吸附，表现出优异吸附性能和去除效率 [5, 8-10]。

**氧化石墨烯的高性能取决于它高比表面的二维结构和表面丰富的各种基团，已成为共识**。通过对鳞片石墨的插层氧化制取氧化石墨，再进一步剥离与分散后制得氧化石墨烯。不同方法制取的氧化石墨烯，表面氧基团种类大多相同，但数量和结构分布上略有不同，主要区别来源于使用的氧化剂体系

不同：Brodie（1895）\ L. Staudenmaier （1898）方法，使用发烟硝酸-高氯酸钾（强芳香环氧化体系）氧化石墨；Hummers（1958）方法，使用高锰酸钾-浓硫酸（$Mn_2O_7$氧化体系）氧化石墨；Jones 方法，使用铬酸-硫酸（不完全氧化体系，常见于膨胀石墨制备，辅以超声工艺完成氧化石墨烯制备）氧化石墨。Hummers 法制备工艺相对简单，制取的石墨烯氧化程度高，表面基团丰富，成为目前材料制备的常用方法，也是商业化生产最常用方法。氧基团的多样化又使氧化石墨烯结构表达变得复杂，目前由 Lerf 和 Klinowski 等人提出的氧化石墨烯非晶结构模型广为科学界接受。他们通过实验总结出氧化石墨烯结构上的基团及分布规律，主要有环氧基（面内）、羧基（边缘）、酮基（分散）、羟基（分散）、烯烃结构等（见图1）[11]。该模型从理论上为氧化石墨烯丰富且独特的物理化学性质解释给予强有力的支撑。含氧基团倾向于与亲水性物质相互作用，而烯烃结构容易与非极性物质或结构相互作用。具体来说，不同种类氧基团在石墨烯表面，可与其他分子或离子间形成静电作用(Electrostatic Interaction)、π-π 交互作用(π-π Bonding)、氢键作用(Hydrogen Bonding)、路易斯酸碱作用(Lewis Acid-base Interaction)、络合作用(Complexation)等[12]，并且通过修饰还可被赋予选择性。因此，从原理上来说，氧化石墨烯作为吸附材料存在着丰富的作用机理。Zhong Hua 等用生物表面活性剂鼠李糖脂对氧化石墨烯的表面羟基进行酯化，制成 RL-GO 复合材料，通过 GO 羧基离子化-静电吸引、π-π 堆栈、氢键等作用，使其对亚甲基蓝（阳离子型染料）的吸附量由 377.6mg/g 提高到 529.1 mg/g，增幅达 28.7%[13]。Chen Wei 等发现自然水体存在的微量 $Na_2S$（5mM）会与氧化石墨烯作用，把环氧基和醚基转化成酚羟基或酮基，提高对 1-萘酚的吸附；表面碳碳疏水结构增多，提高对菲的吸附[14]。水中氧化石墨烯的表面 Zeta 电位可达-30mV 以上，分散很稳定，但若表面基团被还原或损失，表面电位降低易引发石墨烯团聚、沉淀以及表面积的减少。为提高对 RB 染料分子的吸附，需要增加表面疏水性，且为防止团聚，Jae Young Lee 等把氧化石墨烯固定到藻酸水凝胶中，用抗坏血酸水热法转化去除表面基团，使得疏水结构数量增加，吸附性能由 360 mg/g 提高到 730 mg/g[15]。Hui Ying Yang 等则利用 N 掺杂方法，提高氧化石墨烯 N 区域表面羟基活性，增加其对硼酸离子的亲和力，提高在海水中对硼离子的去除率[10]。有意思的是，张建锋等发现在厌氧条件下，水体中 $Na_2S$ 可完全转化还原氧化石墨烯表面羧基，造成铅的解吸附[16]。因此，探究**含氧基团等结构如何变化是推测吸附作用的理论出发点，为基于"氧基团调控"原理利用氧化石墨烯开展环境修复研究提供指导**[17]。

氧化石墨烯的吸附研究中，吸附材料有采用氧化石墨烯分散液、粉末和还原态氧化石墨烯干粉直接作为吸附剂的，也有进行了二维纳米结构表面修饰、负载其他低维材料的氧化石墨烯复合材料作吸附剂的。在内容上普遍参照传统碳材料的研究办法，成熟但缺乏创新。**一直以来，如何更大程度利用**

**氧化石墨烯表面氧基团等结构，使氧化石墨烯具有更高吸附性能和环境修复能力是一个重要内容。**

一般来说，吸附材料表面积越大，吸附位点也越多，相应的吸附性能也越高。由上述可知，氧化石墨烯在水中二维结构几乎完全暴露，理论上活性面积大；但因表面氧基团存在不同种类，性质也不一样。应用于吸附时，针对特定的靶向分子/离子，根据吸附活性大小，这些氧基团可被划分为活性基团/位点和无吸附活性的惰性基团/位点。

当前为了获得石墨烯，溶剂法中对氧化石墨烯表面基团的调控，目的是去除它们，用途多指向电子电气方面的研究，实验条件也比较苛刻，如采用强酸、强碱、高温、高压等条件。但氧化石墨烯若应用到环境修复中，其氧化程度低，氧基团少，可利用范围会变窄，而氧化程度越高，氧基团越丰富，氧基团通过调控转化就可获得更多活性位点，吸附应用范围就大。相关报道中，利用部分/充分还原的氧化石墨烯开展各类吸附已有很多，但依然只利用了氧化石墨烯较少的基团/结构，性能提升依然有巨大空间。因此，**针对不同吸附靶向物，区分氧化石墨烯结构中的吸附活性位点，辨识并活化惰性吸附位点，实现高效吸附，是值得深入研究的方向。**

为强化直接吸附性能，相关研究有提出原位增强吸附（In-situ adsorption enhancement, IAE）方法：初次吸附达到饱和时，对吸附材料进行强化处理，从而与靶向物发生二次吸附，实现吸附材料性能提升。该工作尝试过以吖啶橙吸附去除为例，初步结果表明，氧化石墨烯的吸附容量可由 1900 mg/g 提升到 3300 mg/g，增幅50%左右，优于实验中氧化石墨烯预先被还原再吸附的对照组结果[8]。接着，辅以材料分析发现，水中氧化石墨烯表面有羧基被还原转化成羟基（见图2），导致可氢键作用的活性位点明显增加。目前，考察**在吸附过程中对氧化石墨烯进行原位活化及其对吸附结果影响的研究，鲜有报道。**

氧化石墨烯也被认为是非晶平面大分子/离子材料[18]。制备的氧化石墨烯分子尺寸呈现区间分布，分子的表面基团种类、数量及分布也不统一。这些特点与制备方式密切关系，比如原材料石墨大小、制备过程中的插层氧化时间、第三阶段的水解反应时间、对氧化石墨超声剥离的强度和时间等等[19-22]。同一锅制备的氧化石墨烯，不同尺寸下的石墨烯片，物化性能或各有差异；但一定范围内的氧化石墨烯"分子"物化性质还是比较一致的，类似于纳米石墨烯量子点[23, 24]。而相关研究虽有报道，但缺乏对这方面细致的考察，使吸附机理解释不完整，难以拓展应用，因此有必要进行进一步的深入研究。具体来说，室温下氧化石墨烯在水体吸因羧基转换为羟基，导致吸附量大幅增加，但对**不同尺寸下氧化石墨烯的性能增强程度、过程中因还原导致非极性活性位点的出现（如碳碳烯烃、芳烃结构，无需高温[25]）及其影响、因还原产生形貌变化对吸附的影响[6, 26]，以及还原导致靶向物的解吸附的影响等，未作细究。**而类似的吸附研究可供参考的很少，吸附机理及吸附模型仍需进一

步探究[12]。

围绕以上问题，可继续考察以有机染料吖啶橙为吸附靶向物，结合传统吸附方法与原位光谱、原位显微表征等手段，在单阶段单靶吸附过程之外，提出双阶段单靶吸附、双阶段双靶吸附（见图 3）开展研究，深入考察氧化石墨烯尺寸、形貌、表面基团/结构转化等因素与增强吸附的内在联系，健全、完善原位增强吸附机理，以期充分发挥氧化石墨烯作为高效环境修复材料的优势。该项工作意义在于：一，建立起一种环境材料增强吸附方法，推动石墨烯、碳纳米管等先进碳纳米材料在环境修复领域的研究及加速应用；二，为吸附材料多层次利用及各类新型吸附材料的开发提供方向，抛砖引玉，助力我国水体污染治理和水质保障事业快速发展。

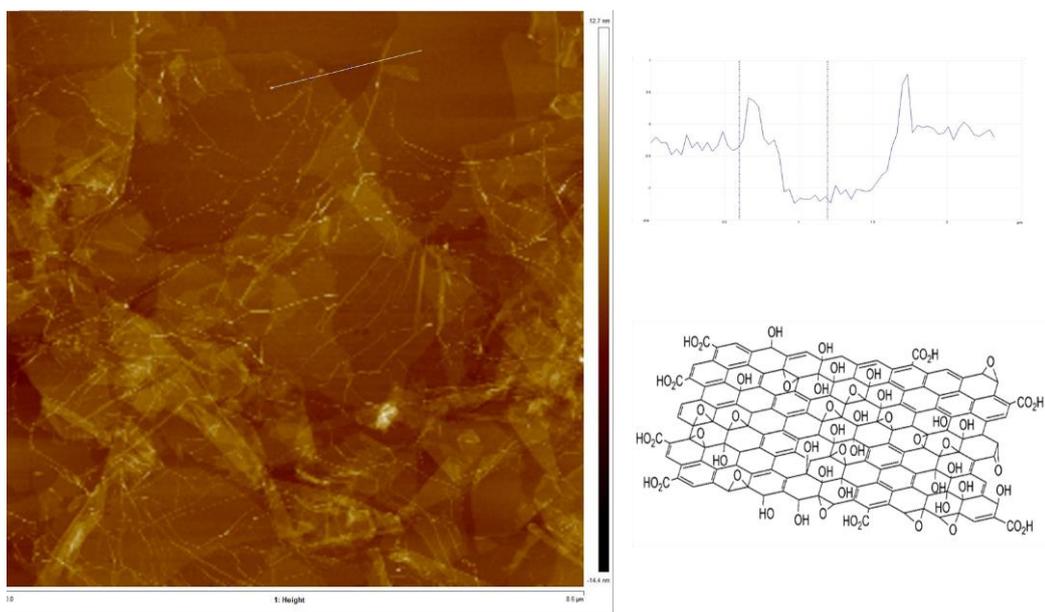

图 1.（左）以膨胀石墨（片层中径~50um）为原材料，利用改进 Hummers 法（2013 年）高效安全制备的氧化石墨烯（产率~99.5%以上）的原子力显微图，样品以片状云母为衬底制成；（右上）厚度曲线，可知石墨烯厚度 ~0.9nm，为单层；（右下）目前最常被引用和接受的用于现象解释的氧化石墨烯结构 L 模型

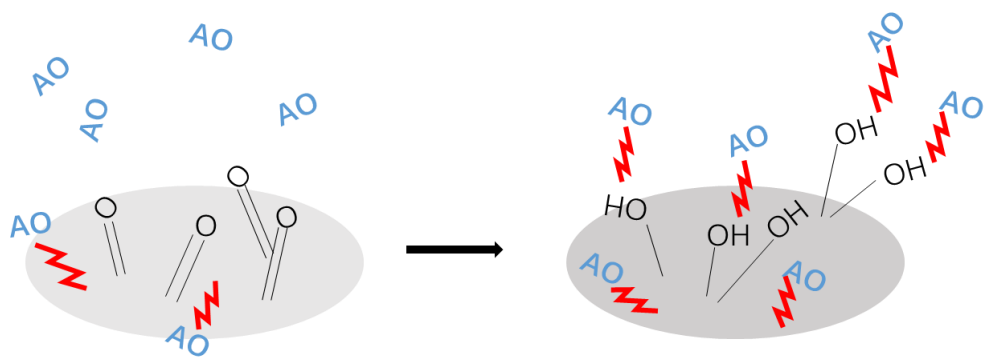

图 2. 双阶段单靶吸附中氧化石墨烯增强吸附吖啶橙机理推测：可发生氢键作用的羟基增多导致原位增强吸附（IAE）

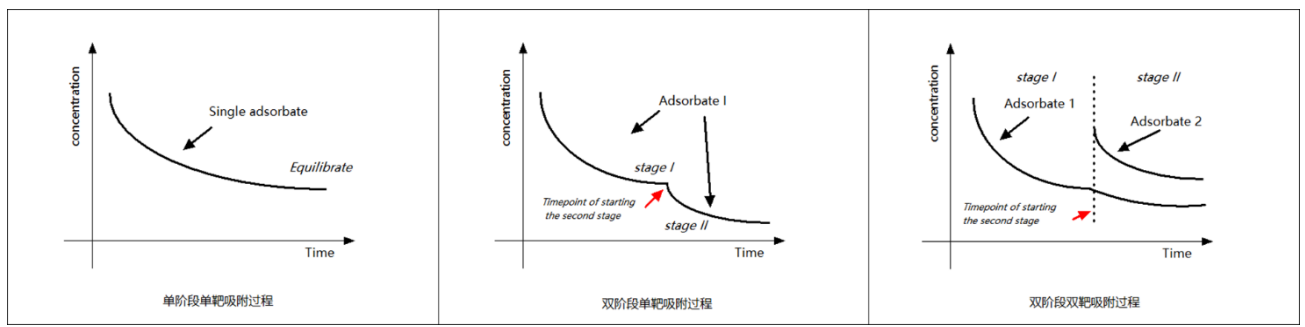

图 3. 三种过程下的靶向物浓度-时间曲线图。（左）"单阶段单靶吸附过程"，即一般吸附过程，靶向物（吸附质）与吸附剂共混吸附，随着材料吸附位点逐渐饱和，溶液浓度不变。（中） "双阶段单靶吸附过程"，为便于研究，特指被研究体系中吸附质（靶向物）浓度降低至平衡时，对吸附剂作原位强化处理，活化材料表面的惰性位点形成新的吸附位点，完成对靶向物的二次吸附，全过程表现为两个吸附平衡阶段。（右）"双阶段双靶吸附过程"，与"双阶段单靶吸附过程"相似，但第一阶段用于对特定活性位点的"惰化"。该阶段中吸附剂与特定靶向物发生作用，吸附平衡后，吸附剂表面的特定位点被占据或消耗而无法继续作用；接着，对吸附材料进行活化，混合添加另一种靶向物后进行双靶向物吸附，直至第二次吸附平衡出现（第一种靶向物浓度变化极小时可不计）。提出该方法主要用于考察不同类型氧基团、结构对吸附的贡献。